\documentclass[12pt,a4paper]{article}
\usepackage[textheight=23cm,textwidth=15cm]{geometry}
\usepackage[utf8]{inputenc}
\usepackage{amsmath}
\usepackage{amssymb}
\usepackage{graphicx}
\usepackage[american]{babel}
\usepackage{here}
\usepackage{color}
\usepackage[articletitle=true, super=false]{achemso}
\usepackage{xcolor}
\usepackage[font=small,labelfont=bf]{caption}
\usepackage{float}

\setcitestyle{square}
\newcommand{\braket}[1]{\langle #1 \rangle}

\newcommand{\ket}[1]{| #1 \rangle}

\begin{document}
\begin{center}

{\LARGE\bf
Tailored Coupled Cluster Theory in Varying Correlation Regimes
}

\vspace{1cm}
{\large
Maximilian M\"orchen$^{a}$\footnote{ORCID: 0000-0002-7467-5719},
Leon Freitag$^{a,b}$\footnote{ORCID: 0000-0002-8302-1354},
and
Markus Reiher$^{a,}$\footnote{Corresponding author; e-mail: markus.reiher@phys.chem.ethz.ch; ORCID: 0000-0002-9508-1565}
}\\[4ex]
$^{a}$ ETH Z\"urich, Laboratorium f\"ur Physikalische Chemie, Vladimir-Prelog-Weg 2, 8093 Z\"urich, Switzerland \\

$^{b}$ Universit\"at Wien, Institut f\"ur theoretische Chemie, W\"ahringer Str. 17,\\ 1090 Vienna, Austria \\

19.11.2020\\
\vspace{.43cm}
\textbf{Abstract}
\end{center}
\vspace*{-.41cm}
{\small
The tailored coupled cluster (TCC) approach is a promising \textit{ansatz} that preserves the simplicity of
single-reference coupled cluster theory, while incorporating a multi-reference wave function through amplitudes
obtained from a preceding multi-configurational calculation.
Here, we present a detailed analysis of the TCC wave function based on model systems, which
require an accurate description of both static and dynamic correlation.
We investigate the reliability of the TCC approach with respect to the exact wave function.
In addition to the error in the electronic energy and standard coupled cluster diagnostics, we exploit the overlap of TCC and full configuration interaction wave functions as a quality measure.
We critically review issues such as the required size of the active space,
size-consistency, symmetry breaking in the wave function, and the dependence of TCC on the reference wave function.
We observe that possible errors caused by symmetry breaking can be mitigated by employing the determinant with the largest weight in the active space as reference for the TCC calculation.
We find the TCC model to be promising in calculations with active orbital spaces which include all orbitals with a large single-orbital entropy, even if the active spaces become very large and then may require modern active-space
approaches that are not restricted to comparatively small numbers of orbitals.
Furthermore, utilizing large active spaces can improve on the TCC wave-function approximation and reduce the size-consistency error, because the presence of highly excited determinants affects the accuracy of the coefficients of low-excited determinants in the active space.
}

\section{Introduction}
\label{sec:introduction}
The solution of the time-independent electronic Schr\"odinger equation is given by an expansion
in a complete basis of Slater determinants (i.e., electronic configurations)
constructed from a finite orbital basis. In this so-called full configuration interaction (FCI)
approach, the expansion coefficients are determined by diagonalization of the Hamiltonian matrix in the complete
determinantal basis.
Since FCI includes every possible electronic configuration within the many-electron Hilbert space constructed of the pre-defined orbital basis, it delivers the exact wave function in this finite basis set, but scales exponentially with the number of orbitals.
The simplest way to achieve polynomial scaling is the truncation of the FCI expansion in a systematic manner, typically defined in terms of a maximum ’excitation’ level, which measures the maximum number of orbital substitutions in some reference determinant. This reference is usually the Hartree-Fock (HF) determinant, in which, in a closed-shell case, the energetically lowest lying canonical HF orbitals are doubly occupied with electrons. \\
In contrast to FCI, which is invariant under a linear transformation of the orbitals, truncated CI depends on the particular choice of reference determinant.
A consequence of the truncation is that truncated CI lacks important features of a reliable wave function
such as size-consistency and size-extensivity \cite{Bartlett2007a}.
By contrast, the coupled cluster (CC) \textit{ansatz} \cite{Coester1958, Coester1960,Cizek1966, Paldus1972a} restores these important properties through its
exponential expansion, even  at a truncated level \cite{Bartlett1981, Taylor1994}.
Additionally, it was recently shown by employing sub-system embedding sub-algebra, that the SRCC \textit{ansatz} can be interpreted as an embedding formalism \cite{Kowalski2018, Baumann2019}.
CC including single and double excitations (CCSD) \cite{Purvis1982a} and perturbative triples CCSD(T) \cite{Raghavachari1989b, Bartlett1990} provides an excellent compromise of accuracy and cost, which is why it is referred to as the \textit{gold standard} in
computational quantum chemistry.\\
It is desirable to include the most important determinants of arbitrary excitation level
in a CI expansion in a systematic and well-defined manner. Since it is not clear which and how many of such determinants
there will be, it is convenient to prune the orbital space to a subspace, which is referred to as the complete active space (CAS), in which FCI is still feasible.
Since quantum chemical approaches for strong electron correlation problems are usually active space methods, they
separate the orbital space into two parts --- a comparatively small one that collects (near-)degenerate and
highly entangled orbitals \cite{Boguslawski2012, Stein2017, stei16, stei16a,stei17}
in the active space and a large one comprising the rest of the virtual orbitals. As a consequence, electron correlation is
divided into two classes: (1) dynamic correlation, which can be recovered by including excited determinants from the entire orbital space \cite{Mok1996}, and (2) static correlation, caused by (near-degenerate) determinants, which have a large weight in the wave function \cite{Hollett2011}.\\
Obviously, active space methods such as CAS configuration interaction (CASCI) and CAS self-consistent field (CASSCF)\cite{Roos1980} are also
constrained by exponential scaling and by a lack of dynamic correlation as no excitations outside of the active space are considered.
The introduction of the density matrix renormalization group (DMRG) \cite{White1992c, White1993}
to quantum chemistry \cite{Legeza2008, Chan2008, Chan2009, Marti2010a, mart11, Chan2011b, Schollwock2011a, Wouters2014a, Kurashige2014, Olivares2015, Szalay2015, Yanai2015, Knecht2016a, Baiardi2020a, Freitag2020}
made considerably larger active spaces accessible due to a decomposition of the CI coefficient tensor that
brings about polynomial scaling.
Another powerful new large-CAS method emerged in the past decade: the FCI quantum Monte Carlo (FCIQMC) approach \cite{Booth2009, Cleland2010}, in which the wave function's coefficient tensor is represented by numbers of \textit{walkers}.\\
Although large active spaces accessible by FCIQMC and DMRG recover an increasing amount of dynamic correlation,
the majority of this type of electron correlation is not accounted for as far more virtual orbitals remain excluded from the active space.
To then recover this remaining dynamic correlation is a major challenge because
all post-diagonalization methods available for that purpose are approximate compared to
the dynamic correlation in the active space evaluated by exact diagonalization. To avoid an
imbalance in the description of dynamic correlation, which could be created when many weakly correlated orbitals are in the
active space, while the majority of the virtual orbitals remains to be considered in a subsequent perturbation-theory step, one is advised to choose a small active
space of all strongly entangled orbitals and leave the dynamic correlation for a consistent separate treatment.\\
By contrast, the standard CC approach depends on a single reference (SR) determinant. Hence, it is not suited for systems with strong static correlation.
In other words, for molecular systems where HF provides a good approximation --- e.g., for the ground state of many organic molecules in their equilibrium structure, SRCC will work accurately.
However, for many chemical problems such as bond-breaking processes, excited states, and open-shell
transition metal  as well as rare earth complexes, SRCC will become inaccurate and may fail \cite{Evangelista2011}.
These types of problems require several determinants to be used as a reference for the many-body expansion of the wave function.\\
A way to combine static and dynamic correlation with polynomial cost are multi-reference (MR) methods.
The most common MR method is MRCI \cite{Buenker1978}, in which, initially, a reference space is defined that
aims to comprise the most important excitations.
From these determinants, a truncated CI expansion is performed which includes excitations into the entire Hilbert space up to the truncation level.
Since MRCI is neither size-consistent nor size-extensive, multi-reference CC (MRCC) approaches have been developed as a way out of these problems \cite{Bartlett2007a, Lyakh2012, Evangelista2018}.\\
Current MRCC approaches can be divided into two main classes: (i) internally contracted (ic) \cite{Banerjee1981} and (ii) Jeziorski-Monkhorst (JM) \cite{Jeziorski1981}.
One issue of both classes was the \textit{multiple-parentage} problem \cite{Evangelista2018}, i.e., the redundant description of amplitudes.
By formulating ic-MRCC as universal valence theory in Fock space \cite{Haque1984, Lindgren1987} and JM-MRCC as state universal theory in Hilbert space \cite{Jeziorski1981, Piecuch1992}, the redundant description could be circumvented.
However, most MRCC formalisms based on the JM \textit{ansatz} suffer from numerical instabilities due to intruder states which are caused by (near-)degenerate determinants \cite{Jankowski1992, Paldus1993, Schucan1972, Schucan1973}.
JM-MRCC theories are plagued by at least one issue that originate from large CASs: (i) for each reference determinant they
require a separate expansion so that the number of parameters grows with the size of the CAS; (ii) it is only orbital-invariant in the
core and in the virtual space, but not in the CAS;
(iii) in truncated JM-MRCC most residuals cannot become zero (except those of a determinant that can be reached from all reference determinants). For these reasons, JM-MRCC theories face increasing difficulties for increasing CAS sizes
(see, e.g., Refs. \cite{Hanrath2008b, Evangelista2018} where state-universal and state-specific MRCC theories, which both belong to JM-MRCC,
are discussed in this respect). \\
In Fock space (FS) MRCC, the intruder state problem could be circumvented by incorporating the intermediate Hamiltonian (IH) \cite{Malrieu1985, Meissner1995} in the FS-MRCC formalism
\cite{Landau1999, Landau2000, Landau2001, Landau2001a, Landau2004, Eliav2005, Musial2005, Musial2011a, Musial2011b, Musial2011, Musial2012, Musial2013, Tomza2013}. It should be noted that this \textit{ansatz} requires the evaluation of so-called sectors, which is computationally very demanding so that it is usually not feasible for large molecular systems.
Implementations of ic-MRCCSD and ic-MRCCSDT without any approximation were presented by Evangelista and Gauss \cite{Evangelista2011a} and by Hanauer and K\"ohn \cite{Hanauer2011}, respectively.
The ic-MRCC method has the advantage of featuring roughly the same number of cluster operators as single-reference CC.
As a consequence, ic-MRCC can, at least in principle, handle larger CAS sizes (containing less than, say, twenty orbitals)
by contrast to JM-MRCC. However,
ic-MRCCSD calculations are expensive as they require elements of the 5-body reduced density matrix
so that only small-CAS calculations are carried out in practice,
which may create a problem for the optimized orbital basis if the active space is overly restricted.
However, the method can be made invariant with respect to orbital rotations in core, active, and virtual spaces and the reference wave function can be optimized self-consistently.
Nevertheless, the cluster operators do not commute since they can create and annihilate electrons in the active space. Hence, the
Baker-Campbell-Hausdorff series does not truncate naturally and one is forced to introduce criteria for its truncation.
Further issues are the average treatment of electron correlation and the redundant description of excitations, which have to be removed by linear transformation.\\
Canonical transformation (CT) theory \cite{Yanai2006, Yanai2007}
corresponds to the class of ic-MRCC theories that is applicable to large active spaces.
The idea behind CT is a cumulant expansion of operators \cite{Evangelista2012}.
However, this cumulant expansion will not be exact in the limit of an infinite expansion
due to an unbalanced treatment of some of the terms,
which is a major problem in CT theory.
Another drawback of this method is the slow convergence of the expansion so that many recursive commutator
iterations are required which makes the overall computational cost very difficult to predict. \\
In view of the issues MRCC approaches suffer from, the most common MR methods capable of describing both static and dynamic
correlation are still multi-reference configuration singles ('S') and doubles ('D')
with a Davidson correction ('+Q'), i.e., MRCISD+Q \cite{Buenker1978, Langhoff1974, Meissner1988a} and multi-reference perturbation theory (MRPT) \cite{Andersson1990, Angeli2001, Angeli2001a}, which both come also with significant drawbacks and immense costs (all require elements of higher-order
density matrices --- up to 4- and 5-body terms).
MRCISD+Q uses a truncated CI expansion so that the resulting wave function lacks the same properties that truncated CI itself lacks.
MRPT is a perturbation \textit{ansatz} and requires large basis sets to describe the electron cusp properly; typically it poorly describes long-range dynamic correlation.\\
A conceptually simpler alternative to \textit{genuine} MRCC methods are MR-driven SRCC approaches introduced by Oliphant and Adamowicz \cite{Oliphant1992}, by Paldus and collaborators \cite{Li1997, Peris1998, Planelles1994, Paldus1994, Peris1997}, and by Stolarczyk \cite{Stolarczy1994}, where an SRCC wave function is externally corrected by another MR wave function. These approaches are based on a single reference only, but include excitations from an active space, a procedure that introduces static correlation effects into an SRCC wave function. \\
Specialized CC variants have been proposed that may alleviate some problems of single-reference CC. For instance, the completely renormalized CC (CR-CC) theory is a special case of the method of moments CC (MM-CC) approach, where the perturbative correction is renormalized by a functional which scales with fully described CC amplitudes \cite{Piecuch2004, Kowalski2000, Kowalski2000a} so that it produces high accuracy for the ground state and
even more accurate energies in the dissociation limit than CCSD(T).
Especially methods based on the CC(P;Q) formalism \cite{Baumann2017, Shen2012, Shen2012a, Shen2012b} in combination with FCIQMC can recover high-order CC energies by requiring only a fraction of the highly excited amplitudes \cite{Deustua2017}.\\
Another MR-driven SRCC \textit{ansatz} introduced by Kinoshita, Hino, and Bartlett is Tailored CC (TCC) \cite{Kinoshita2005}.
In its singles and doubles variant, TCCSD, CI coefficients for singly and doubly excited determinants are
first extracted from a CAS wave function and then transformed to CC amplitudes.
The CAS state parameters included in this way in an SRCC wave function are kept frozen so that they \textit{tailor}
the remaining CC amplitudes during their optimization.
TCCSD was used by Hino {\it et al.} to calculate the equilibrium structure of ozone and its vibrational frequencies \cite{Hino2006}.
Later, Lyakh, Lotrich, and Bartlett also included perturbative triples into the TCCSD wave function for the calculation of the automerization of cyclobutadiene \cite{Lyakh2011}.
Since TCC prevents relaxation of the CAS coefficients, they are decoupled from the CC amplitudes.
To improve on the CAS coefficients, Melnichuk and Bartlett used an extended active space to include more orbitals in the optimization of the CAS \cite{Melnichuk2012, Melnichuk2014}.
Naturally, this way of letting the state parameters responsible for static correlation see more of the dynamic correlation
points toward CAS approaches that allow for truly large active spaces, i.e., FCIQMC and DMRG.
Veis {\it et al.} were the first to employ DMRG to tailor the TCC amplitudes \cite{Veis2016}.
Later, Faulstich {\it et al.} provided a mathematical analysis of this TCC-DMRG approach in which
they formally assessed its advantages and disadvantages \cite{Faulstich2019c}.
Subsequently, these authors presented a numerical DMRG-TCC study at the example of dinitrogen dissociation, which focused on the DMRG in TCCSD\cite{Faulstich2019b}. Recently, Vitale \textit{et al.} published a benchmark on a TCC based on FCIQMC \cite{Vitale2020} and the distinguishable cluster approach \cite{Kats2013, Kats2014, Kats2015, Kats2016, Kats2017}.\\
In view of the salient properties of TCC and the fact that the above-mentioned analyses have not considered an analysis of the
CC part of the wave function or a comparison with FCI, we intend to close this gap in the present paper.
We will examine the TCCSD method at the example of well-known model systems that require static and dynamic correlation.
Due to the small size of these systems, FCI calculations are possible, so that we have access to the best possible wave function.
Since the energy is not necessarily the best diagnostic for the suitability of a wave-function approximation,
we also consider the overlap of the TCC wave function with the exact FCI reference wave function.
Moreover, T$_1$ and D$_1$ diagnostics \cite{Lee1989, Lee1989a, Janssen1998} are employed to measure the quality of the reference Slater determinant.

\section{Theory}
\label{sec:theory}
\subsection{Single-Reference Coupled Cluster}
The SRCC wave function is defined by
\begin{equation}
  \ket{\Psi_{\text{CC}}} = e^{\hat{T}} \ket{\Phi_0},
\end{equation}
where the cluster operator $\hat{T}$ denotes to a linear combination of cluster operators $\hat{T}_\nu$ of excitation level $\nu$.
The exponential can be Taylor expanded and the truncated SRCC is then size-consistent and size-extensive \cite{Bartlett1981, Taylor1994}.
Another advantage of this \textit{ansatz} is the faster convergence to the FCI limit than truncated CI --- despite the fact
that the amplitude equations are not solved variationally but by projection. \\
The cluster operator is defined as
\begin{equation}
  \hat{T}_\nu = \frac{1}{(\nu !)^2} \sum_{\substack{a_1, \dots, a_\nu\\ i_1, \dots, i_\nu}} t_{i_1, \dots, i_\nu}^{a_1, \dots, a_\nu}
  \hat{a}_{a_1}^\dagger \dots \hat{a}_{a_\nu}^\dagger \hat{a}_{i_1} \dots \hat{a}_{i_\nu},
\end{equation}
where $a_p^\dagger$ and $a_q$ are creation and annihilation operators and $ t_{i_1, \dots, i_\nu}^{a_1, \dots, a_\nu}$ represents the amplitude for the corresponding excitation.
The indices $p,q,r,...$ represent generic orbitals, while $i,j,k,...$ and $a,b,c,...$ correspond to hole and particle indices, respectively.
The amplitudes in SRCC are optimized by the amplitude equation
\begin{equation}
  0 = \braket{\Phi_\nu | e^{-\hat{T}} \hat{H} e^{\hat{T}} \Phi_0},
\end{equation}
whereby an excited determinant $\Phi_\nu$ is projected onto the similarity transformed Hamiltonian $e^{-\hat{T}} \hat{H} e^{\hat{T}}$ acting on the reference determinant $\Phi_0$.
Note that we deliberately write the reference wave function in ket notation to highlight the non-hermiticity of the similarity transformed Hamiltonian.
Because of the similarity transformation, the Hamiltonian is no longer Hermitian.
The advantage of this formulation is the natural truncation of the Baker-Campbell-Hausdorff series after the 4-fold commutator.
With the optimized amplitudes, the SRCC energy can be evaluated as follows
\begin{equation}
  \braket{\Phi_0 | e^{-\hat{T}} \hat{H} e^{\hat{T}} \Phi_0} = E_{\text{CC}}.
\end{equation}

\subsection{Tailored Coupled Cluster}
TCC combines an active space method with SRCC leading to a formulation that is supposed to describe both static and dynamic correlation of a system \cite{Kinoshita2005}.
Since the coefficients from an active space are not optimized during the SRCC process, TCC corresponds to the class of externally corrected CC theories.
The TCC wave function employs the split-amplitude \textit{ansatz} \cite{Piecuch1993, Piecuch1994, Jankowski1996}, which means that the linear expansion
of the cluster operators $\hat{T}_\nu$ is divided into operators that act either on the active ('CAS') or on the external ('ext') space,
\begin{equation}
  \ket{\Psi_{\text{TCC}}} = e^{\hat{T}} \ket{\Phi_0} =  e^{(\hat{T}^{\text{ext}} + \hat{T}^{\text{CAS}})} \ket{\Phi_0} = e^{\hat{T}^{\text{ext}}} e^{\hat{T}^{\text{CAS}}} \ket{\Phi_0}.
\end{equation}
Because $\hat{T}^{\text{ext}}$ and $\hat{T}^{\text{CAS}}$ never include the same set of orbitals
and act on a single reference determinant, the operators commute with each other \cite{Kinoshita2005}.
In the TCCSD variant, CCSD is corrected by the singly and doubly excited CAS amplitudes.

The excitation of the CAS coefficients is defined with respect to a reference determinant, which must be the same as for the CC expansion.

Therefore, the TCCSD wave function
\begin{equation}
  \ket{\Psi_{\text{TCCSD}}} = e^{(\hat{T}^{\text{ext}}_{1} + \hat{T}^{\text{ext}}_{2})} e^{(\hat{T}^{\text{CAS}}_{1} + \hat{T}^{\text{CAS}}_{2})} \ket{\Phi_0}
\end{equation}
contains the single and double excitations from both spaces, implying that the coefficients of
higher excitations in the CAS are neglected in the TCC approach.
Hence, TCCSD includes the same excitations as (SR)CCSD, where the CAS amplitudes are frozen during the optimization of the external amplitudes \cite{Kinoshita2005}. Moreover,
the TCCSD wave function has the same number of parameters as the analogous SRCCSD wave funcion so that the computational scaling is comparable.
Compared to a \textit{genuine} MRCC, the TCC approach therefore has the advantage that problems, which are related to
multiple reference determinants such as the \textit{multi-parentage} problem, do not occur.
If the active space amplitudes were allowed to relax, they would converge to the corresponding CCSD amplitudes and would have no corrective effect on the external amplitudes. \\
The CI coefficients from the CAS calculation, which correspond to single and double excitations, must be transformed into CC amplitudes based on the relationship between cluster, $\hat{T}$,  and CI, $\hat{C}_i$, operators
\begin{align}
  \begin{split}
  \hat{T}^{\text{CAS}}_{1} & = \hat{C}_{1}, \\
  \hat{T}^{\text{CAS}}_{2} & = \hat{C}_{2} - \frac{1}{2} \left[ \hat{C}_{1} \right]^{2}.
\end{split}
\end{align}
In the case of a  CI wave function that is not intermediate-normalized, i.e. $\braket{\Psi_{\text{CAS}} | \Phi_{\text{HF}}} = 1$,
the transformation to CC amplitudes
requires normalization by the coefficient $c_0$, which corresponds to the reference determinant:
\begin{align}
  \begin{split}
  t_{i}^{a} & = \frac{c_{i}^{a}}{c_0}, \text{\quad $a,i \in $ CAS,} \\
  t_{ij}^{ab} & = \frac{c_{ij}^{ab}}{c_0} - \frac{(c_{i}^{a} c_{j}^{b} - c_{i}^{b} c_{j}^{a})}{c_0^2}, \text{\quad $a,b,i,j \in $ CAS}.
  \label{eq:transformation}
\end{split}
\end{align}
Because the exact ('full') CC energy depends only on single and double amplitudes (calculated in the presence of
all higher-excited amplitudes from the amplitude equations) and as we calculate the single and double amplitudes from an FCI in a CAS, i.e.,
also under the condition that they are exact from a full CC point of view, we can calculate the CAS energy with
a CCSD {\it ansatz} exploiting the amplitudes obtained from the CASCI calculation.
Hence, the CAS energy is obtained after the transformation of the coefficients by
 \begin{equation}
  \braket{\Phi_0 | \hat{H} e^{\hat{T}^{\text{CAS}}} \Phi_0}_c = E_{\text{CAS}},
  \label{eq:tcc:casenergy}
\end{equation}
where $c$ means that only connected terms are involved \cite{Kinoshita2005}.
Since the CAS amplitudes replace the corresponding CCSD amplitudes, the external cluster operators must be redefined as
\begin{align}
  \begin{split}
  \hat{T}_1^{\text{ext}} & = \sum_{a,i} t_{i}^{a} \hat{a}_a^\dagger \hat{a}_i, \text{\quad $\{a,i\} \not\subset $ CAS,} \\
  \hat{T}_2^{\text{ext}} & = \frac{1}{4}\sum_{a,b,i,j} t_{ij}^{ab}  \hat{a}_a^\dagger \hat{a}_b^\dagger \hat{a}_j \hat{a}_i, \text{\quad $\{a,b,i,j\} \not\subset $ CAS}
\end{split}
\end{align}
to prevent external excitation within the CAS space, meaning that the whole set of indices is not allowed to be element of the CAS. Hence, if for example $a,b,i$ address orbital indices within the CAS, but $j$ does not, the whole set of indices is not element of the CAS and therefore part of the external space.
This implies that excitations from the external occupied to the active virtual space and from the active occupied into the external virtual space are part of the external amplitudes.
Hence, the amplitude equations of the TCCSD formalism are defined as
\begin{align}
  \begin{split}
  & \braket{\Phi_{i}^{a} | \hat{H} e^{(\hat{T}^{\text{ext}}_{1} + \hat{T}^{\text{ext}}_{2})} e^{(\hat{T}^{\text{CAS}}_{1} + \hat{T}^{\text{CAS}}_{2})} \Phi_0}_c = 0, \text{\quad $\{a,i\} \not\subset $ CAS},  \\
  & \braket{\Phi_{ij}^{ab} | \hat{H} e^{(\hat{T}^{\text{ext}}_{1} + \hat{T}^{\text{ext}}_{2})} e^{(\hat{T}^{\text{CAS}}_{1} + \hat{T}^{\text{CAS}}_{2})} \Phi_0}_c = 0,
  \text{\quad $\{a,b,i,j\} \not\subset $ CAS},
\end{split}
\end{align}
Only the external space amplitudes are optimized and the pre-calculated CAS amplitudes encode the static correlation in the system.
The correlation energy \cite{Lowdin1959} is then obtained as a sum of the active space energy $E_{\text{CAS}}$ and the remaining external energy $E_{\text{ext}}$ by
\begin{equation}
  \braket{\Phi_0 | \hat{H} e^{(\hat{T}^{\text{ext}}_{1} + \hat{T}^{\text{ext}}_{2})} e^{(\hat{T}^{\text{CAS}}_{1} + \hat{T}^{\text{CAS}}_{2})} \Phi_0}_c = E_{\text{CAS}} + E_{\text{ext}} = E_{\text{TCCSD}}.
\end{equation}

As it has been shown by Lyakh {\it et al.} \cite{Lyakh2011}, TCCSD can be supplemented with a perturbative triples correction.
For its evaluation it is crucial to exclude every triple-excited amplitude which depends purely on CAS singles and doubles. \\
In this work, we consider an analysis of TCCSD without any perturbative correction, because (i) this has been the standard TCC model so far and (ii) it is not plagued by additional computational burden (noting though that such a correction may still be decisive for a truly reliable
TCC approach). We will turn to triples corrections in the TCC framework in future work.

\subsection{Diagnostic Descriptors}

\subsubsection{$c_0$ Coefficient}
As a first estimation of the MR character of a wave function, the CI coefficient $c_0$, which corresponds to the reference determinant, can be taken into account.
If the HF determinant is a good approximation of the system, the $c_0$ coefficient should be close to one for a normalized wave function. \\ 

\subsubsection{T$_1$ Norm}
Therefore, for SRCC wave functions the T$_1$ and D$_1$ diagnostics \cite{Lee1989, Lee1989a, Janssen1998}, which are independent of the $c_0$ coefficient, may be used to assess the reliability of a CI or CC wave function.
The weighted T$_1$ norm is defined as
\begin{equation}
  \text{T}_1 = \sqrt{\frac{\vec{t}_1 \cdot \vec{t}_1}{N}},
\end{equation}
where the vector $\vec{t}_1$ contains all single amplitudes and $N$ represents the number of electrons.
The number of single-excitation CC amplitudes increases with the number of electrons $N$.
Hence, the Euclidean norm of the amplitudes is divided by $\sqrt{N}$ to make T$_1$-diagnostics independent of the number of electrons.
The threshold value for T$_1$ that points to a reliable CC wave function is 0.02 \cite{Lee1989a}.

\subsubsection{D$_1$ Norm}
Owing to the sensitivity of T$_1$ diagnostics to orbital rotations, the D$_1$ norm was introduced as
\begin{equation}
  \text{D}_1 = \sqrt{\lambda_{\text{max}}\left( \mathbf{T}\mathbf{T}^\dagger \right)},
\end{equation}
where $\mathbf{T}$ represents the single-excitation amplitude matrix, where the columns and rows correspond to holes and particles, respectively. $\lambda_{\text{max}}$ is the largest eigenvalue of the matrix $\mathbf{T}\mathbf{T}^\dagger$.
The threshold value of  D$_1$ to classify a CC wave function as reliable is 0.05 \cite{Janssen1998}.
By virtue of the diagonalization of $\mathbf{T}\mathbf{T}^\dagger$, the D$_1$ norm is less sensitive to orbital rotations.

\subsubsection{Energy and Overlap Error}
The most reliable assessment of an approximate theory is a comparison with FCI in the same orbital basis.
Since the CC equations are solved by projection, we define an energy error in terms of FCI for a CC wave function indicated by the subscript $x$ as
\begin{equation}
  \Delta E = | \braket{ \Psi_{\text{FCI}} | \hat{H} | \Psi_{\text{FCI}} } - \braket{ \Phi_{\text{0}} | \hat{H} | \Psi_{x} } |.
\end{equation}
A second option for a direct comparison is one based on the wave function itself, e.g. in terms of
the norm suggested by Kutzelnigg\cite{Kutzelnigg1991}
\begin{equation}
  ||\Psi_{x} - \Psi_{\text{FCI}}||^2 = 2 \left( 1 - \braket{ \Psi_{\text{FCI}} | \Psi_{x} } \right).
\end{equation}
The norm vanishes when $\Psi_x$ becomes exact.
In this work, we exploit the overlap error
\begin{equation}
  \Delta_{\text{ov.}} = | 1 - \braket{ \Psi_{\text{FCI}} | \Psi_{x}} |
\end{equation}
as a measure for the accuracy of a wave function.
Obviously, the prohibitively steep scaling of FCI prevents us from exploiting energy and overlap errors
in routine calculations as standard diagnostics.

\section{Results\label{sec:results}}
For the investigation of the TCCSD wave function we chose the model systems P4 \cite{Jankowski1980} and H8 \cite{Jankowski1985} to analyze the behavior of the method in the statically and dynamically correlated regimes.
We compare all TCSSD results with results obtained with its single-reference analog, SRCCSD (for the sake of brevity, we drop the prefix 'SR' in the following),
and with FCI data as the exact result within the given orbital basis set.
Because the wave function is more sensitive to errors than the energy,
T$_1$ and D$_1$ diagnostics \cite{Lee1989, Lee1989a, Janssen1998} and especially the overlap of wave functions are
in the focus of our analysis.

Since TCCSD corresponds to the class of MR-driven SRCC methods, the wave function still depends on a single reference determinant.
To assess the effect of the reference determinant on the quality of the TCCSD wave function, we performed all calculations twice, based on the two most dominant determinants.

Nevertheless, the orbital basis also greatly affects the wave function.
Since TCCSD requires the choice of an active space, we evaluated every property in the HF and CAS optimized orbital basis
for the sake of comparison.

\subsection{Computational Methodology}
All calculations were carried out with the PySCF program package \cite{Sun2018}.
This includes the calculation of integrals, the HF and CAS orbital optimization, as well as the CASCI, FCI and CCSD calculations.
To ensure that the same orbital basis and CI expansion is used throughout this work, we interfaced our in-house TCCSD implementation with PySCF.
For all calculations on the P4 and H8 model, we followed Jankowski {\it et al.} \cite{Jankowski1980, Jankowski1985} and chose
DZP $(4s1p/2s1p)$ \cite{CanalNeto2005} and DZ $(4s/2s)$ basis sets \cite{Dunning1970, Dunning1977}, respectively.
The entanglement analysis in section \ref{subsec:h8_cas88} was performed with the AutoCAS program
\cite{stei16,stei17,Stein2019} that allows one to identify the orbitals responsible for the largest static correlation. \cite{stei16,Boguslawski2012}. Such analysis is based on entanglement entropy measures\cite{Legeza2003a, Rissler2006, Legeza2006}
such as the one-orbital (or single-orbital) entropy
\begin{equation}
  s_p(1) = - \sum_{\beta = 1}^4 w_{\beta, p} \ln w_{\beta, p}
  \label{eq:singleOrbitalEntropy}
\end{equation}
calculated from the eigenvalues $w_{\beta, p}$ of the one-orbital reduced density matrix for orbital $p$, where $\beta$ is
an one of the four possible configurations for a spatial orbital, i.e., unoccupied, $\alpha$- or $\beta$-occupied, or doubly occupied. It can be obtained from elements of the one- and two-particle reduced density matrix \cite{Boguslawski2015}.
The mutual information of two orbitals $p$ and $q$
\begin{equation}
  I_{pq} = \frac{1}{2} \left[ s_p(1) + s_q(1) - s_{pq}(2) \right] (1 - \delta_{pq})
  \label{eq:mutualInformation}
\end{equation}
requires the corresponding two-orbital entropy
\begin{equation}
  s_{pq}(2) = - \sum_{\beta = 1}^{16} w_{\beta, pq} \ln w_{\beta, pq},
\end{equation}
where $w_{\beta, pq}$ are the eigenvalues of the two-orbital reduced density matrix and
$\beta$ now refers to all 16 possible basis states (occupation combinations) of two spatial orbitals.

\subsection{P4 Model}
\begin{figure}[h!t]
  \centering
  \includegraphics[width=.15\textwidth]{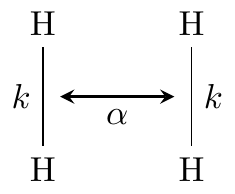}
  \caption{The P4 model in which $\alpha$ is a parameter that is characteristic for the distance between the two H$_2$ units and $k$ is constant at 2\,bohr.}
  \label{fig:p4_model}
\end{figure}

The P4 model \cite{Jankowski1980} consists of two hydrogen molecules with a constant intramolecular distance of 2\,bohr.
The two fragments are aligned parallel to each other, resulting in a planar rectangular configuration with the distance between the two molecules determined by the parameter $\alpha$ (see Figure \ref{fig:p4_model}).
The P4 model has a compressed configuration for $\alpha < 2$, a square configuration for $\alpha = 2$, and a stretched configuration for $\alpha > 2$.

The stretched and compressed configurations have $D_{2h}$ symmetry with the respective dominant determinants
in short-hand notation (see supporting information for the complete set of HF orbitals)
\begin{align}
\label{P4_1}
  \ket{\Phi_1} & = \ket{(1a_{1})^2(1b_{2u})^2}, \\
\label{P4_2}
  \ket{\Phi_2} & = \ket{(1a_{1})^2(1b_{3u})^2},
\end{align}
while the quadratic configuration represents a degenerate case.
The active space is defined by CAS(2\,e$^{-}$, $(1b_{2u})(1b_{3u})$).
The P4 model covers both SR and MR regimes.
The MR character of the wave function can be tuned by the parameter $\alpha$.
At large separation $\alpha$ of the two molecules, size-consistency can be probed.\\

\begin{figure}[h!t]
  \includegraphics[width=\textwidth]{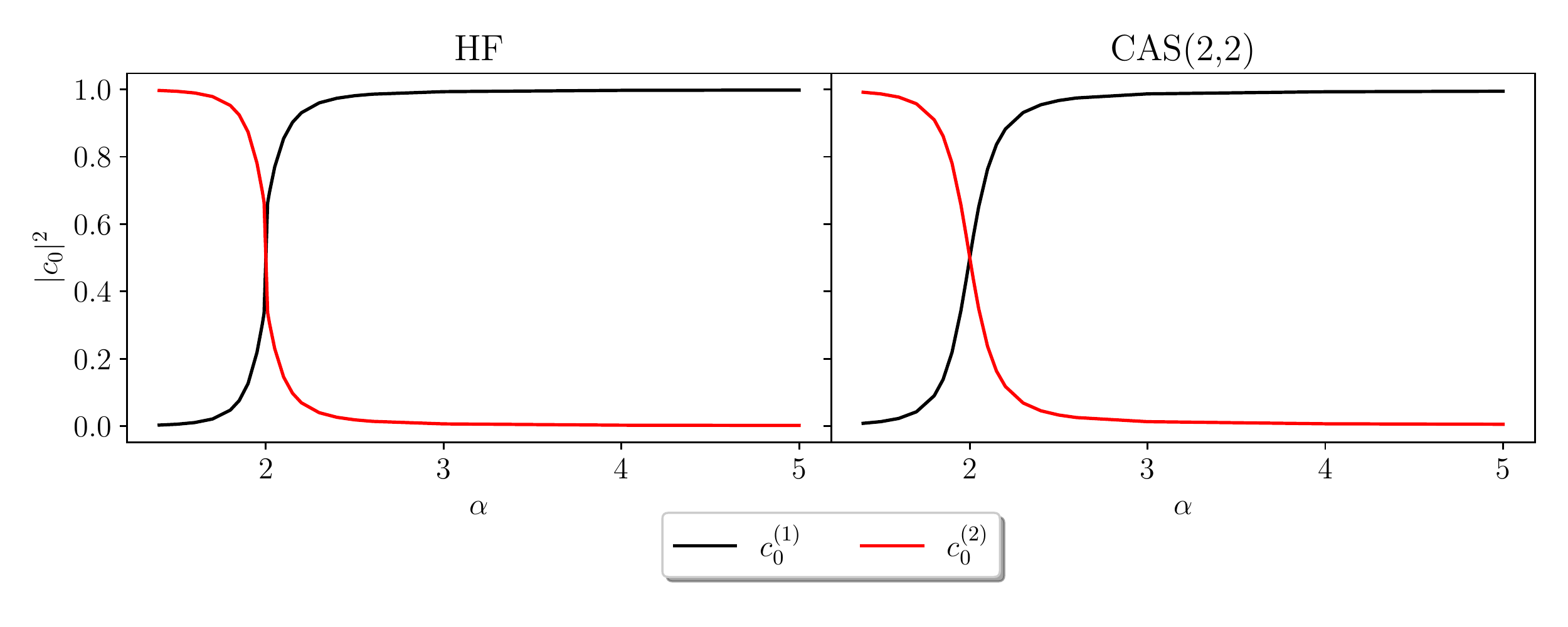}
  \caption{Square of the P4-model CI coefficient ($|c_0|^2$) of $\Phi_1$ ($c_0^{(1)}$) and $\Phi_2$ ($c_0^{(2)}$) of Eqs.
(\ref{P4_1}) and (\ref{P4_2})
for different values of $\alpha$.
Both expansion coefficients were obtained for determinants constructed from different molecular orbital bases: (1) HF
orbitals (left) and (2) CAS(2,2) orbitals (right).
}
  \label{referenceCoeff_p4}
\end{figure}

The importance of the individual determinants is evident from Figure \ref{referenceCoeff_p4}.
It becomes also clear that the use of CAS(2,2) orbitals, compared to HF orbitals, leads to no significant change of the $c_0$ coefficient for the corresponding reference determinant.

$\ket{\Phi_2}$ represents the most dominant reference for the compressed configuration and $\ket{\Phi_1}$ the one for the stretched configuration.
Between $1.75 < \alpha < 2.25$, the wave function shows a significant multi-configurational character. These two
 determinants become completely degenerate at $\alpha = 2$.
We denote determinants as $\Phi_i$ and wave functions which depend on a reference $i$ as $\Psi_i$.
\begin{figure}[h!t]
  \includegraphics[width=\textwidth]{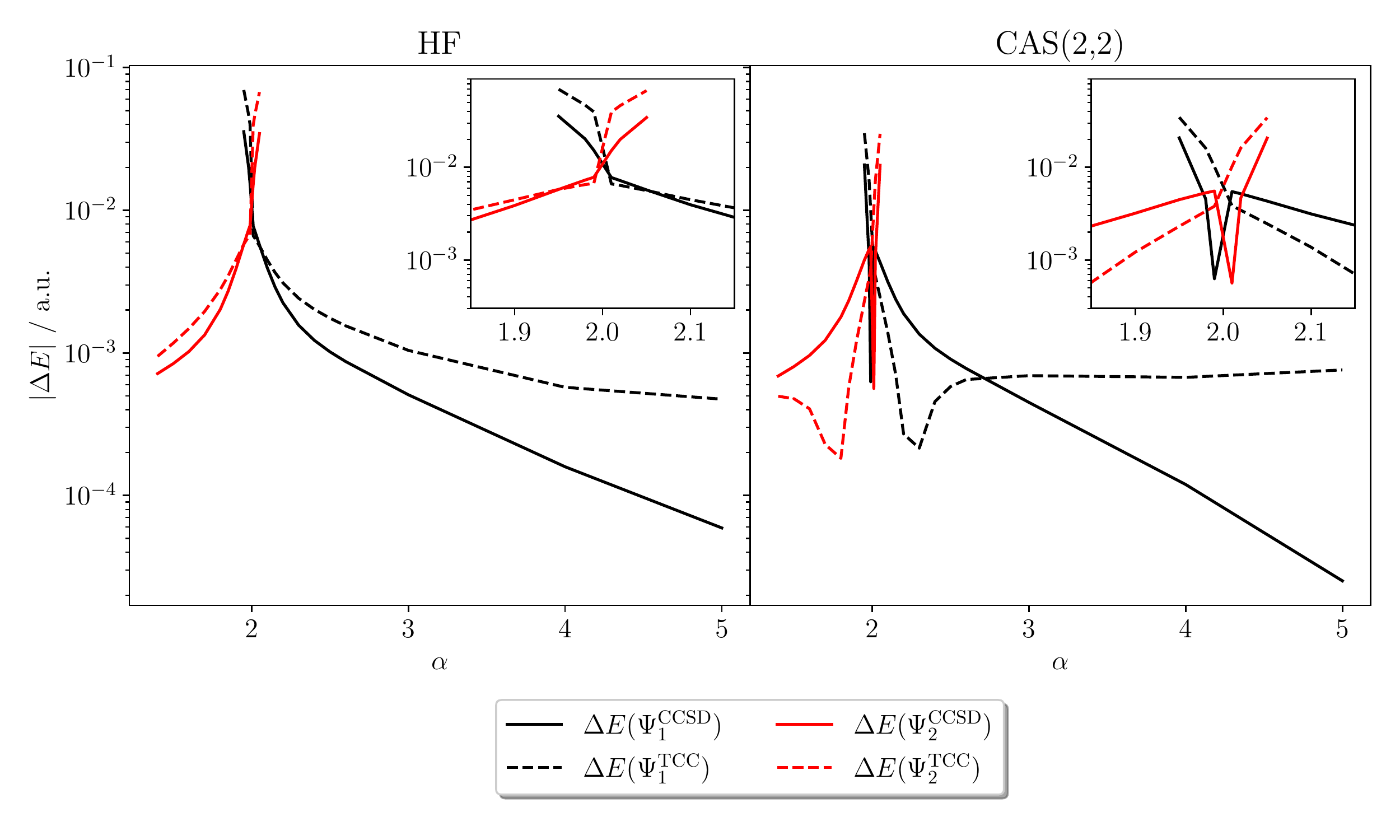}
  \caption{Absolute energy error $|\Delta E|$ of the P4 model for TCCSD(2,2) and CCSD with respect to the FCI energy based on the reference determinant $\Phi_1$ and $\Phi_2$ and in different orbital bases: (1) HF orbitals (left) and CAS(2,2) orbitals (right).}
  \label{E_error_p4}
\end{figure}
In the following, we study the reliability of the TCCSD and CCSD methods for different values of $\alpha$.
We expect the reliability of CCSD to deteriorate as the determinant that was not chosen as the reference for CCSD
starts to dominate in the wave function, while TCCSD should be able to describe the MR region qualitatively correct.\\
Figure \ref{E_error_p4} depicts the energy profile given as the absolute error with respect to the FCI energy
(see supporting information for the potential energy surfaces).
From this figure it is obvious that TCCSD is not a \textit{genuine} MR method, since it shows a strong dependence on the reference, even if the most important orbitals are both included in the active space.
Another problem is the energy error of the TCCSD in the SR region ($\alpha < 1.75$ and $\alpha > 2.25$), because it is still larger than the corresponding error for CCSD, when HF orbitals are used.
Especially for the dissociated configuration ($\alpha >$ 4.5), the difference between TCCSD and CCSD increases, because TCCSD will, in contrast to CCSD, in general not be size-consistent \cite{Veis2016}.
TCCSD is size-consistent only in special cases when the active space can be
divided into a direct product of active spaces of the sub-systems \cite{Kinoshita2005}.
Only for the degenerate case, TCCSD leads to marginally better energies than CCSD.
The reason for this is the qualitatively correct description by the amplitudes obtained from the active space.
Since these amplitudes must not relax, they result in an improved energy compared to CCSD, however the total energy error is still unacceptable for an accurate MR description. \\
This situation changes in the CAS(2,2) optimized orbital basis, where TCCSD still shows a strong dependence on the dominant reference, but for the almost degenerate ones the energy error is smaller than for CCSD.
Since the absolute error is shown and TCCSD is not variational, the decrease near the degenerate configuration is based on a sign change, which also explains the rapid increase of the error around $\alpha \sim 2$.
For the dissociated configuration, due to size-consistency, the CCSD energy approaches FCI, while the error is nearly constant for TCCSD.

\begin{figure}[h!t]
  \centering
  \includegraphics[width=\textwidth]{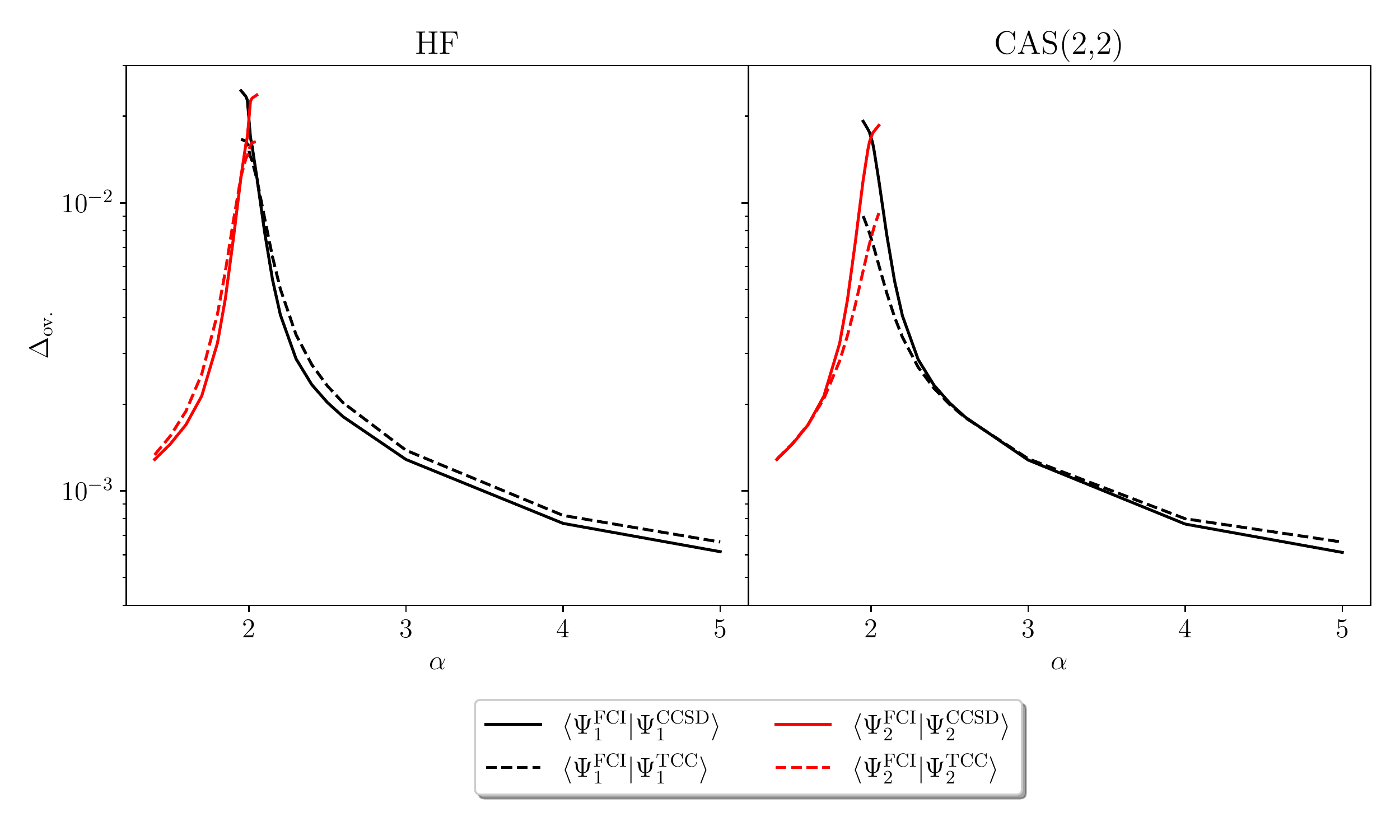}
  \caption{Overlap error $\Delta_{\text{ov.}}$ in the P4 model for CCSD and TCCSD(2,2) with respect to FCI based on the reference determinants $\Phi_1$ and $\Phi_2$ and in different orbital bases: (1) HF orbitals (left) and CAS(2,2) orbitals (right).}
  \label{overlap_p4}
\end{figure}
\begin{figure}[h!t]
  \centering
  \includegraphics[width=\textwidth]{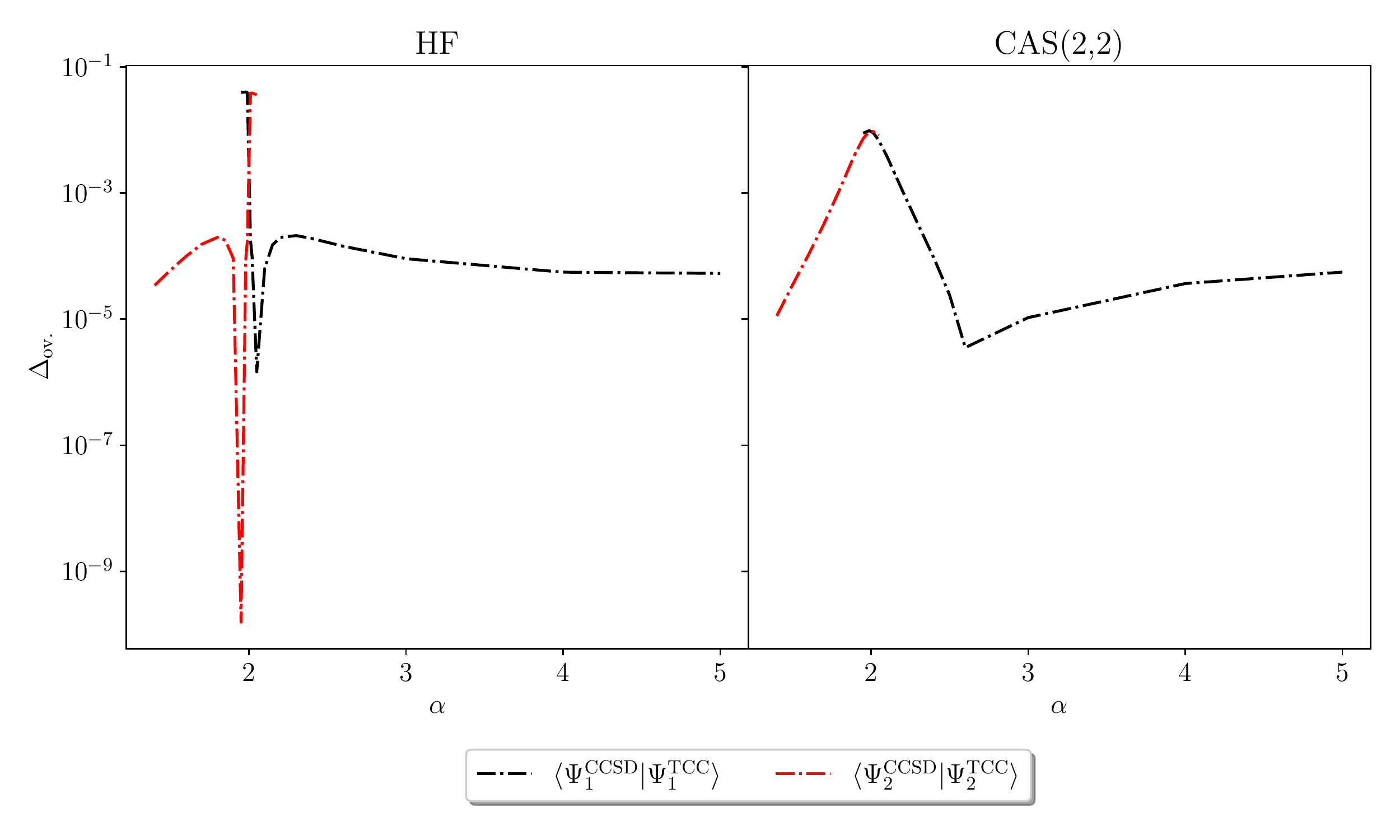}
  \caption{Overlap error $\Delta_{\text{ov.}}$ in the P4 model of CCSD with respect to TCCSD(2,2) based on the reference determinants $\Phi_1$ and $\Phi_2$  and in different orbital bases: (1) HF orbitals (left) and CAS(2,2) orbitals (right).}
  \label{overlap_p4_2}
\end{figure}
As already mentioned, the energy is not necessarily a benchmark for the quality of the wave function.
A straightforward diagnostic for the quality of a wave function is its overlap with the FCI wave function.
Figures \ref{overlap_p4} and \ref{overlap_p4_2} show the overlap error $\Delta_{\text{ov.}}$ of the CCSD and TCCSD wave function with the FCI wave function and with each other to highlight the reliability and differences between the two wave functions.
For HF orbitals, both methods show a small overlap error in the SR-dominated region for $\alpha >3$.
However, the overlap error increase for configurations, where the other determinant becomes more important.
CCSD shows a similar behavior in the CASSCF(2,2) basis, while the overlap error for TCCSD(2,2) is significantly smaller for configurations with a larger MR character than in the HF basis. \\
In the HF orbital basis, the CCSD wave function possesses a somewhat larger overlap with the FCI wave function compared to the TCCSD wave function.
The largest difference between TCCSD and CCSD is again around $\alpha=2$.
The reason is the same as for the smaller energy error for the nearly degenerate configurations, when compared to CCSD.
However, in TCCSD the external amplitudes are based on only one reference, which explains the increase of the overlap error for the degenerate configuration.\\
From the overlap between the CCSD and TCCSD wave function, it can be concluded that the wave functions for the SR-dominated cases differ only slightly, since the external amplitudes are based on the same reference as the CCSD and the active space amplitudes differ not as much from the corresponding ones in CCSD as for the MR region. \\ 
The same applies to the calculations of the SR regions based on CAS(2,2)-optimized orbitals.
The reason for the small overlap error is that the orbitals and active space amplitudes are optimized for the degenerate case, and these contribute most to the overlap between TCCSD and FCI.
\begin{figure}[h!t]
  \includegraphics[width=\textwidth]{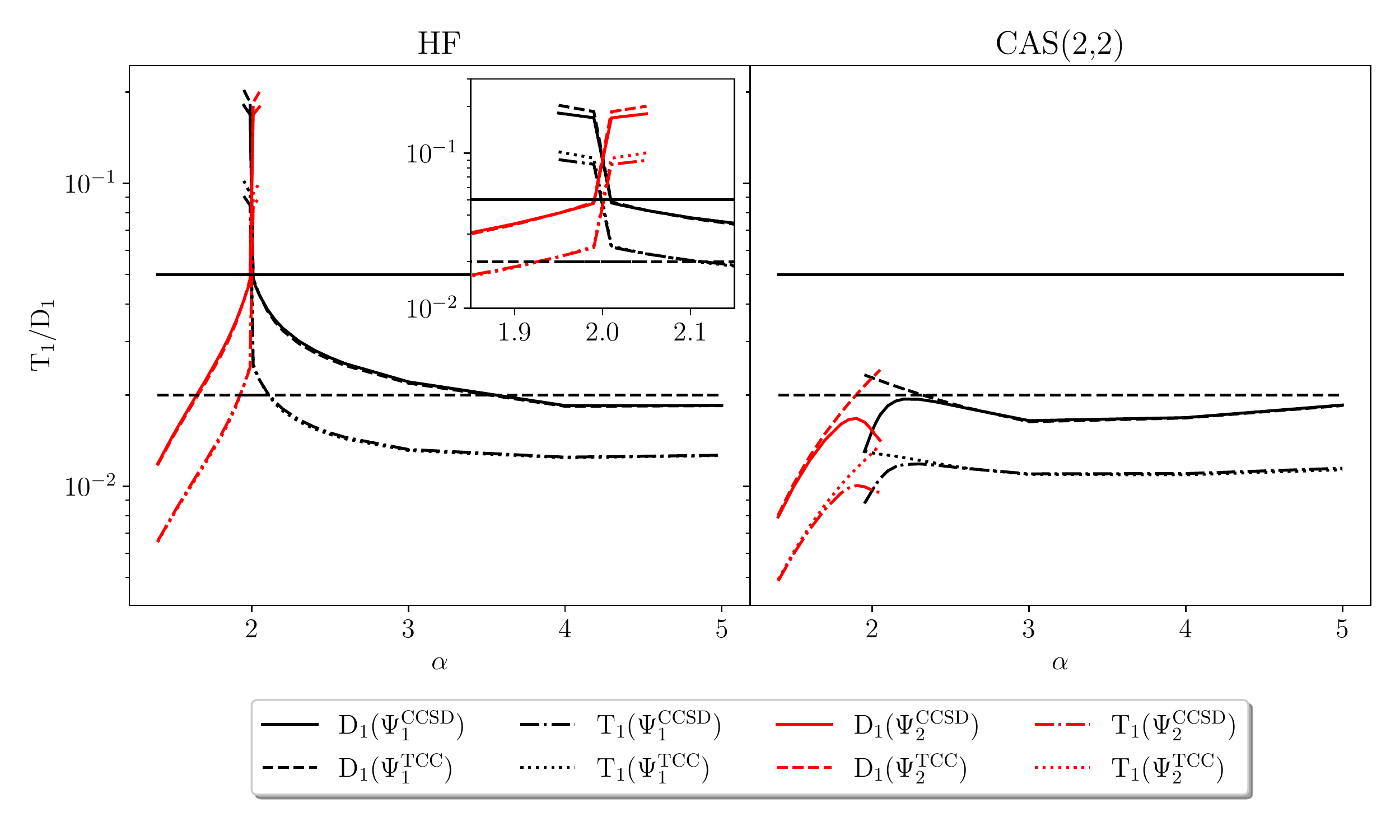}
  \caption{T$_1$ and D$_1$ diagnostics for the TCCSD(2,2) and CCSD P4-model wave functions based on $\Phi_1$ (black lines) and $\Phi_2$ (red lines) and in different orbital bases: (1) HF orbitals (left) and CAS(2,2) orbitals (right). The solid upper line represents the D$_1$ reliability threshold, while the dotted line  corresponds to the T$_1$ reliability threshold.}
  \label{Diagnostics_p4}
\end{figure}
For reasons of completeness also the T$_1$ and D$_1$ diagnostics are shown in Figure \ref{Diagnostics_p4}.
The T$_1$ and D$_1$ diagnostics mostly agree with the overlap diagnostics for SR-dominated regions.
However, the T$_1$ and D$_1$ diagnostics below their thresholds in the range $1.75 < \alpha < 2.25$ contradict the more
reliable FCI overlap error, as they seem to predict
that CCSD and TCCSD deliver a reliable wave function representation in that range, while in fact they do not.
Since the overlap error with respect to FCI provides more reliable diagnostics than the T$_1$ and D$_1$ values based on CASSCF orbitals, we only consider the overlap for the H8 model (see supporting information for T$_1$ and D$_1$ for H8).
It should be noted, however, that T$_1$ and D$_1$ come with no additional computation overhead and are therefore a
practical means to evaluate the quality of general SRCC wave functions.

\subsection{H8 Model with a Small CAS}

\begin{figure}[h!t]
  \centering
\includegraphics[width=.5\textwidth]{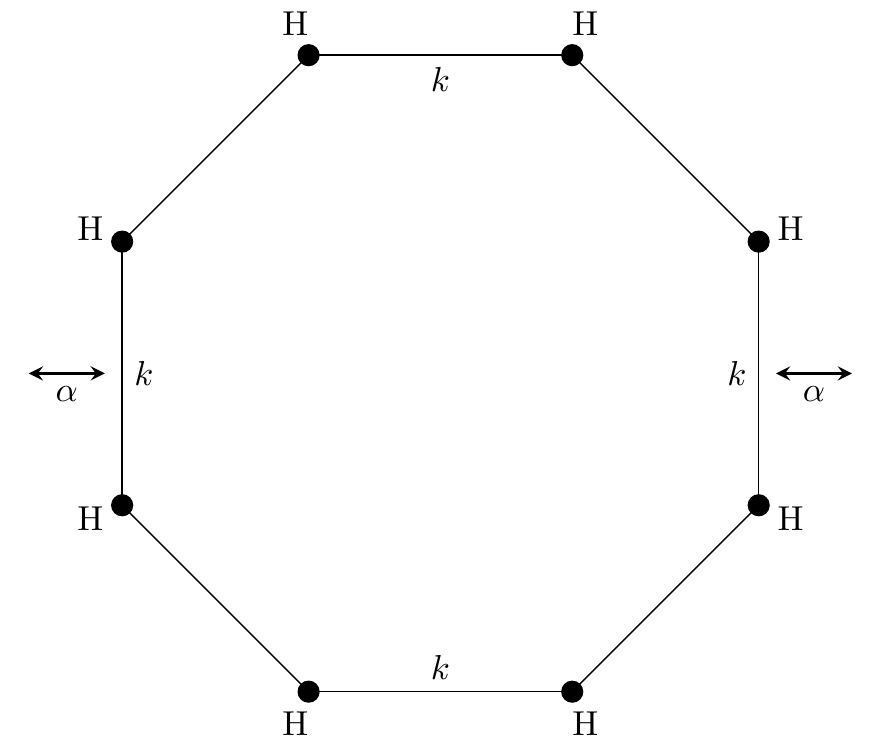}
\caption{The H8 model consists of four stretched H$_2$ units with a constant bond length $k = $2\,bohr. Two units lie on the top and bottom edges of the ocatagon and remain fixed in their position. The distance between the vertically arranged units can be controlled by the parameter $\alpha$ which is characteristic for the distortion of the regular octagonal geometry.}
\label{fig:h8_model}
\end{figure}

The H8 model system \cite{Jankowski1985} consists of four hydrogen molecules with a constant intramolecular distance of 2\,bohr.
The molecules are arranged in an octagonal configuration with $D_{8h}$ symmetry.
Two H$_2$ units which lie on the top and bottom edge of the octagon remain fixed in their position.
The distance between the units at the left and right edge can be controlled by the parameter $\alpha$,
(see Figure \ref{fig:h8_model}).

The variation of $\alpha$ results in a compressed configuration for $\alpha < 0$, an octagonal configuration for $\alpha = 0$ and a stretched configuration for $\alpha > 0$.
The stretched and compressed configuration corresponds to the $D_{2h}$ symmetry, with
\begin{align}
  \label{H8_1}
  \ket{\Phi_1} &= \ket{(1a_{g})^2 (1b_{3u})^2 (1b_{2u})^2 (2a_{g})^2}, \\ 
  \label{H8_2}
  \ket{\Phi_2} &= \ket{(1a_{g})^2 (1b_{3u})^2 (1b_{2u})^2 (1b_{1g})^2},
\end{align}
as the dominant reference determinants in short-hand notation (see supporting information for the complete set of HF orbitals).
To attain a consistent description of the electronic ground state along the deformation of the H8 system, the active space is chosen to be CAS(2\,e$^{-}$, $(2a_{g})(1b_{1g})$).
The octagonal configuration at $\alpha = 0$ represents the completely symmetric, i.e., $D_{8h}$ configuration.
\begin{figure}[h!t]
  \includegraphics[width=\textwidth]{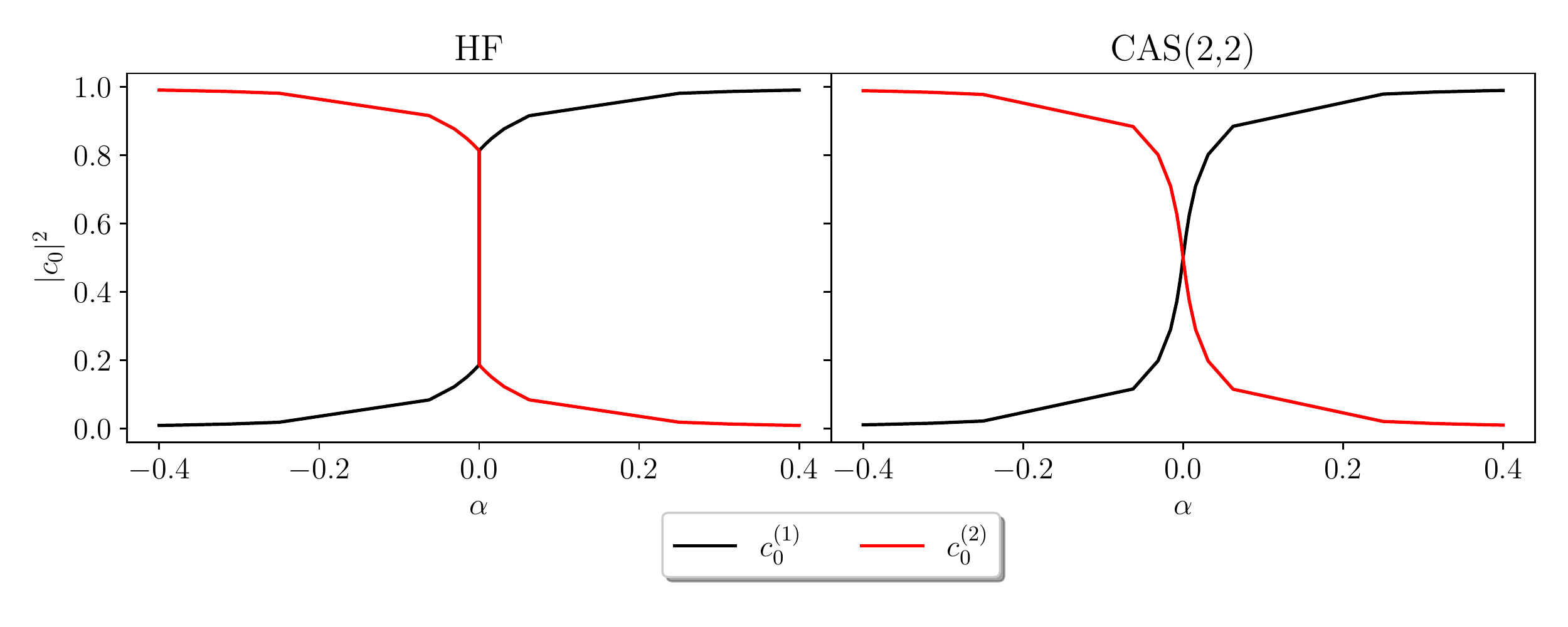}
  \caption{Square of the H8-model CI coefficient ($|c_0|^2$) of $\Phi_1$ ($c_0^{(1)}$) and $\Phi_2$ ($c_0^{(2)}$) of Eqs. (\ref{H8_1}) and (\ref{H8_2}) for different values of $\alpha$.
  Both expansion coefficients were obtained for determinants constructed from different molecular orbital bases: (1) HF orbitals (left) and (2) CAS(2,2) orbitals (right).}
  \label{referenceCoeff_h8}
\end{figure}
Figure \ref{referenceCoeff_h8} shows the square of the $c_0$ CI expansion coefficients for the two determinants.
$\ket{\Phi_2}$ represents the dominant determinant for $\alpha < -0.1$ and $\ket{\Phi_1}$ for $\alpha > 0.1$, while the weight of the determinants increases for $-0.1 < \alpha < 0.1$ and becomes fully degenerate for $\alpha = 0$.
\begin{figure}[h!t]
  \includegraphics[width=\textwidth]{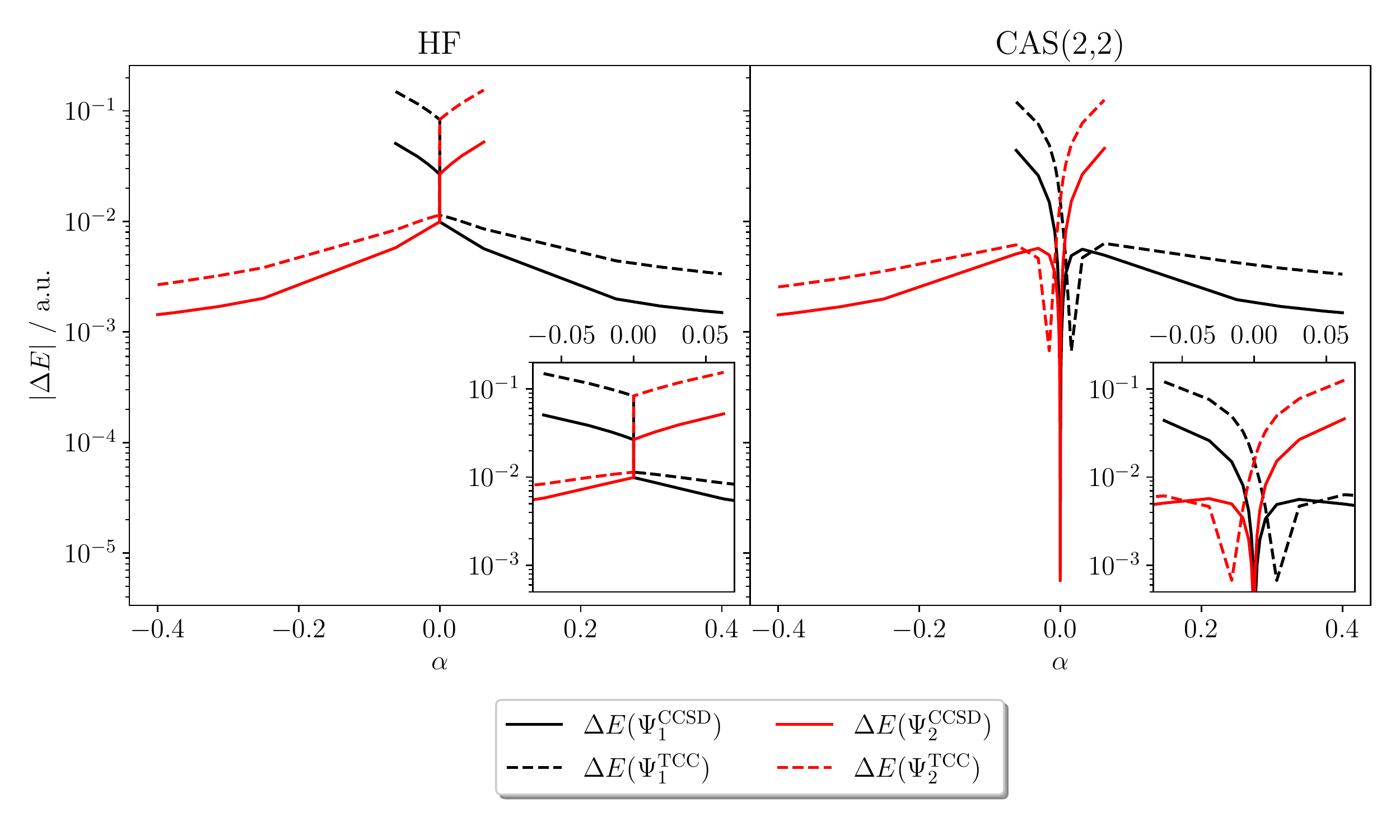}
  \caption{Absolute energy error $|\Delta E|$ of the H8 model for TCCSD(2,2) and CCSD with respect to the FCI energy based on the reference determinant $\Phi_1$ and $\Phi_2$ and in different orbital bases: (1) HF orbitals (left) and CAS(2,2) orbitals (right).}
  \label{E_error_h8}
\end{figure}
Figure \ref{E_error_h8} shows the energy error for TCCSD(2,2) and CCSD with respect to FCI based on HF and CAS(2,2) orbitals.
In the HF orbital basis, CCSD shows a much smaller energy error than TCCSD, in contrast to the energy error profile based on the CASSCF orbitals, where this is only valid only for $\alpha < -0.05$ and $\alpha > 0.05$.
Owing to the absolute value of the error, the spikes near $\alpha=0$ represent a sign change in the error.
Similar to the P4 model, the energy error increases with an increasing MR character of the wave function, but it is surprising that the TCCSD energy error for the high-symmetry configurations
near $\alpha = 0$, which results in a strong MR character, is larger than for CCSD.\\
Because of the optimized CI coefficients in the CASSCF basis, TCCSD(2,2) yields a smaller energy error than CCSD for the near
high-symmetry configurations near $\alpha = 0$.
However, the effect of the amplitudes obtained from the active space is not enough to \textit{tailor} the external amplitudes to describe the nearly degenerate configurations accurately.
Because of the fixed CAS amplitudes, TCCSD shows a larger energy error than CCSD for the SR-dominated regions.
\begin{figure}[h!t]
  \includegraphics[width=\textwidth]{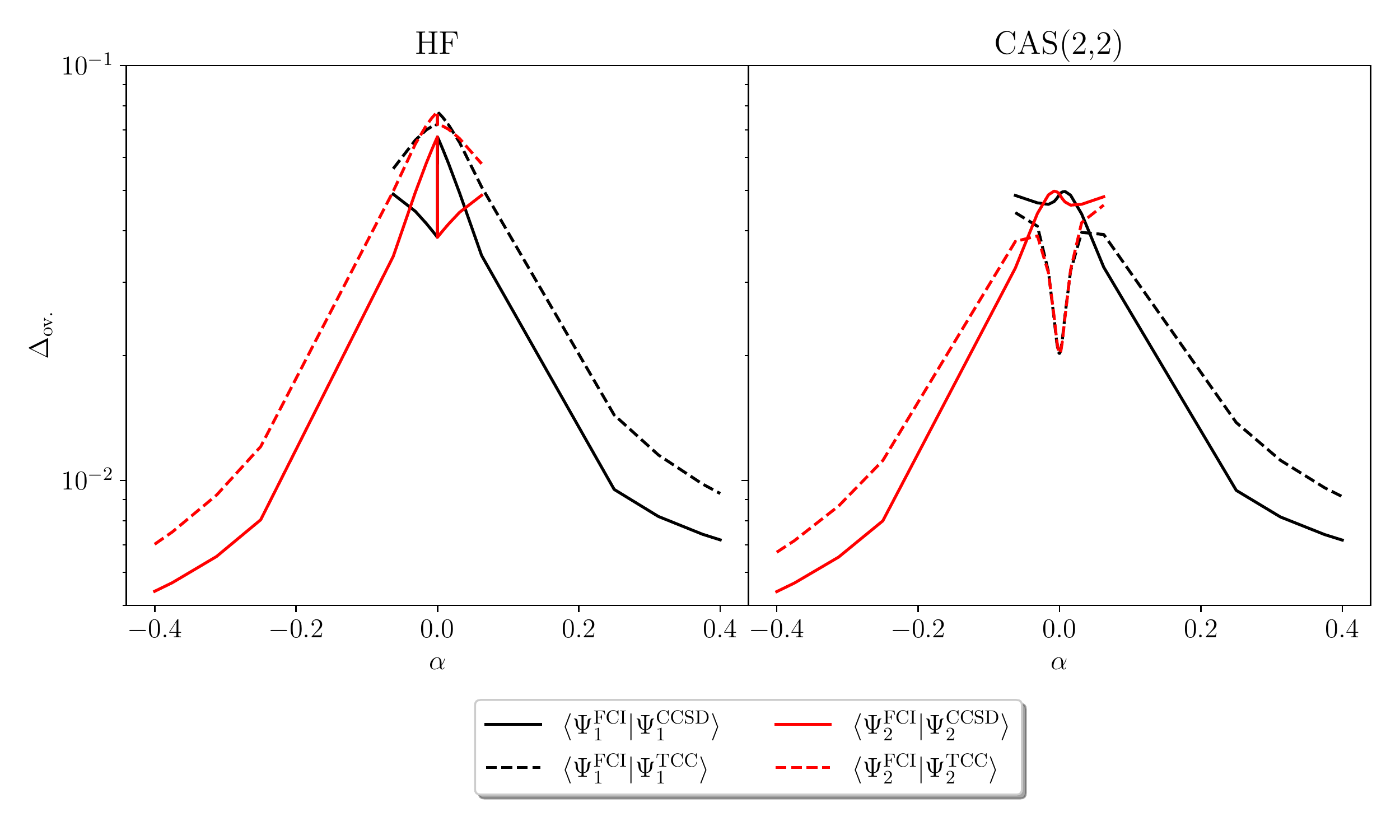}
  \caption{Overlap error $\Delta_{\text{ov.}}$ in the H8 model for CCSD and TCCSD(2,2) with respect to FCI based on the reference determinants $\Phi_1$ and $\Phi_2$ and in different orbital bases: (1) HF orbitals (left) and CAS(2,2) orbitals (right).}
  \label{overlap_h8}
\end{figure}
This is supported by the measurements of the overlap error in Figure \ref{overlap_h8}.
The HF orbital-based CCSD wave function shows for all values of $\alpha$ a smaller overlap error than the TCCSD wave function.
In the CASSCF orbital basis TCCSD yields a smaller overlap error for $-0.05 < \alpha < 0.05$.
Also the TCCSD overlap errors of both TCCSD wave functions diverge only after this region.
However, the overlap error for the MR region is rather large for the CCSD and TCCSD wave functions compared to what we observed for  P4 model and to the errors seen in the SR region. \\
From the $c_0$ coefficients we can conclude, that the H8 model relies more on a MR description than the P4 model.
In view of the overlap error it is clear that TCCSD requires a CAS optimized orbital basis in order to yield a smaller overlap error than CCSD for the MR region around $\alpha = 0$.

\subsection{H8 Model with a Large CAS\label{subsec:h8_cas88}}
\begin{figure}[h!t]
  \includegraphics[width=\textwidth]{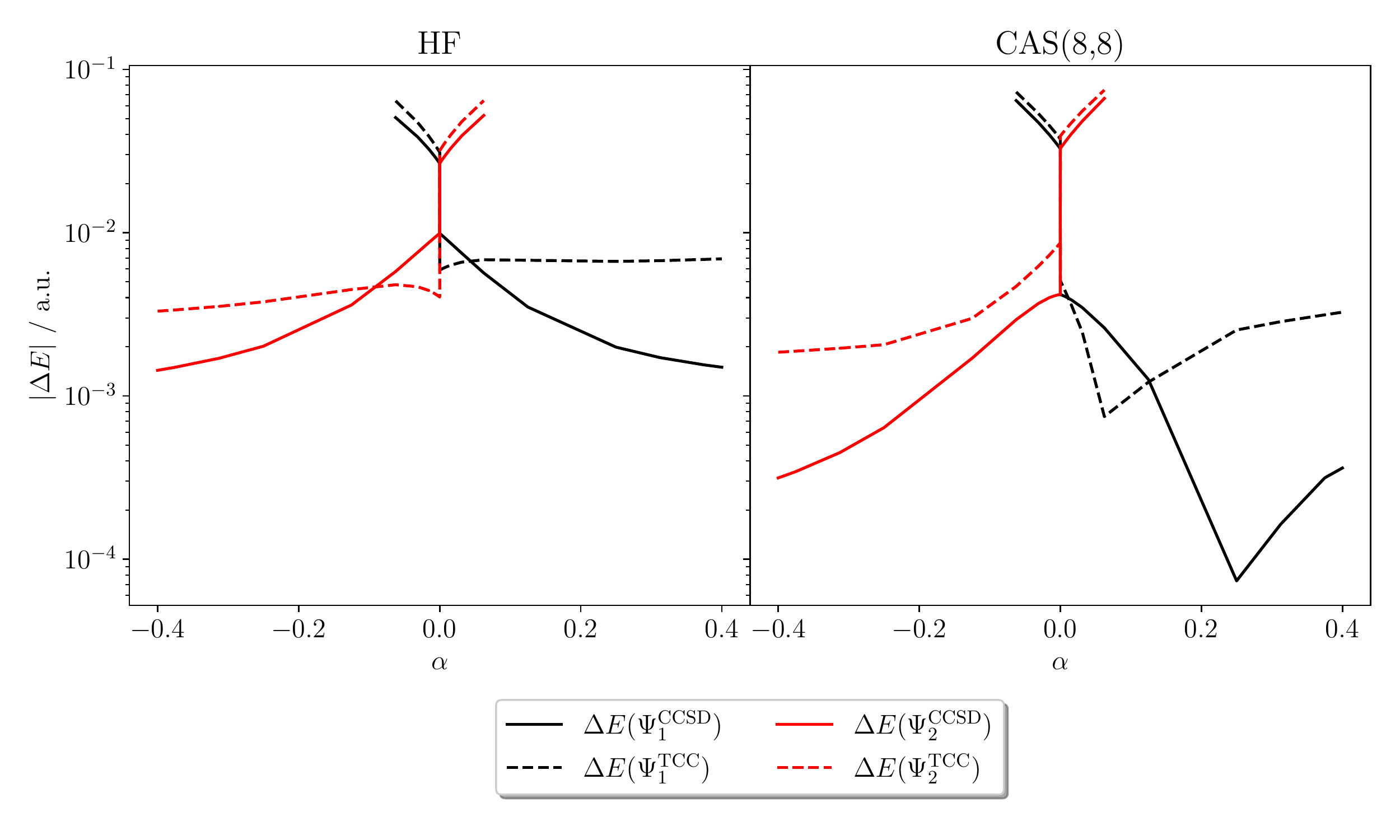}
  \caption{Absolute energy error $|\Delta E|$ of the H8 model for TCCSD(8,8) and CCSD with respect to the FCI energy based on the reference determinant $\Phi_1$ and $\Phi_2$ and in different orbital bases: (1) HF orbitals (left) and CAS(8,8) orbitals (right).}
  \label{E_error_h8_ex}
\end{figure}

\begin{figure}[h!t]
  \includegraphics[width=\textwidth]{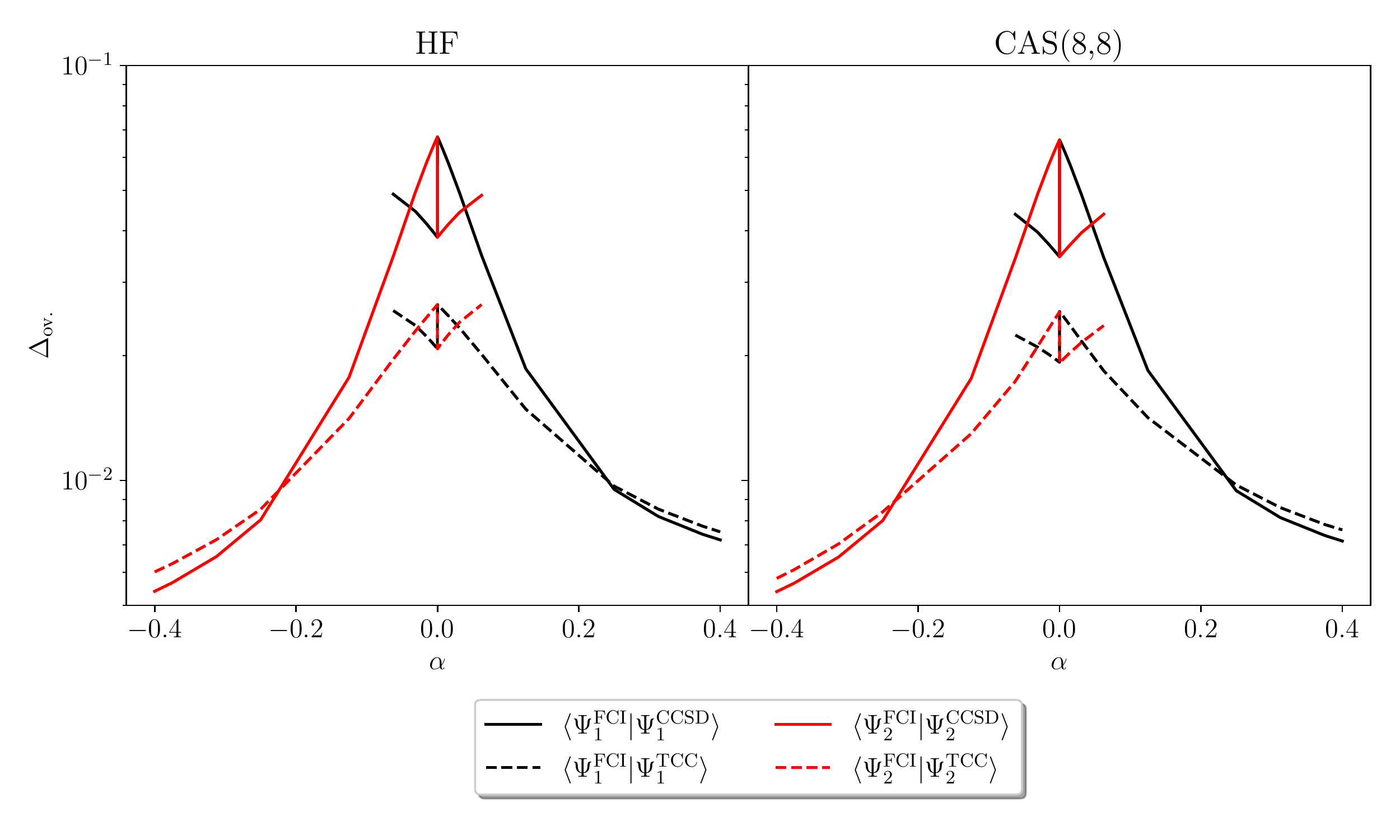}
  \caption{Overlap error $\Delta_{\text{ov.}}$ in the H8 model for CCSD and TCCSD(8,8) with respect to FCI based on the reference determinants $\Phi_1$ and $\Phi_2$ and in different orbital bases: (1) HF orbitals (left) and CAS(8,8) orbitals (right).}
  \label{Overlap_h8_ex}
\end{figure}

Faulstich {\it et al.} emphasized \cite{Faulstich2019c, Faulstich2019b} that in TCC the active space must include all determinants that are mainly responsible for static correlation.
A side effect of the large active space is also a partial reduction of the size-consistency error introduced by the truncation of the active space \cite{Veis2016}.
\begin{figure}[h!t]
  \centering
  \includegraphics[width=.5\textwidth]{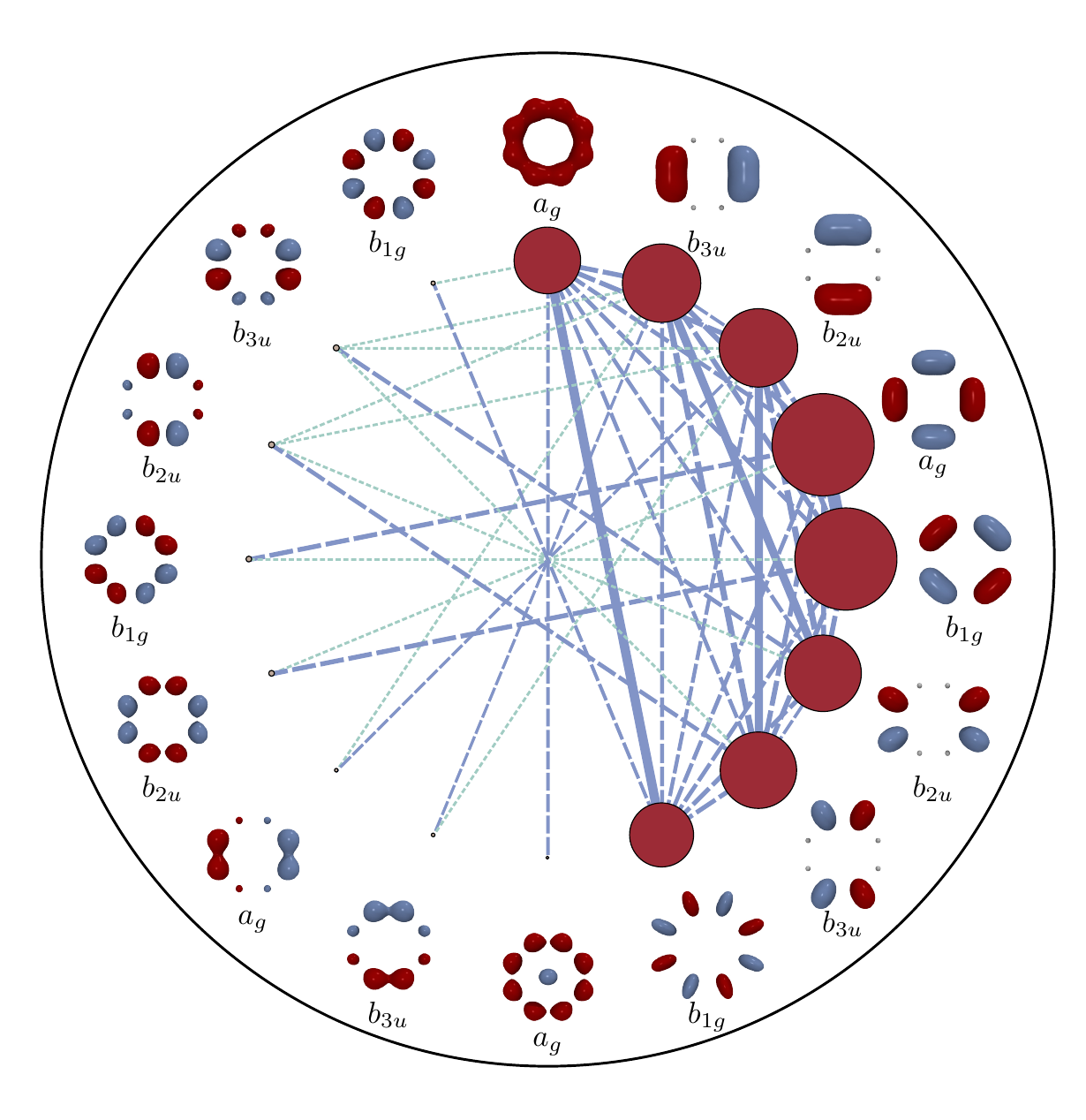}
  \caption{Entanglement diagram for the H8 model at $\alpha = 0$ in the HF orbital basis. The size of the circles represents the single-orbital entropy defined in Eq. \eqref{eq:singleOrbitalEntropy}. The lines between the circles denote the mutual information of Eq. \eqref{eq:mutualInformation} with the thickness representing its magnitude.}
  \label{figOrbEntanglement}
\end{figure}
The orbital entanglement diagram in Figure \ref{figOrbEntanglement} features large single-orbital entropies for eight orbitals and strong mutual information for these orbital pairs. Since static correlation emerges if determinants with a large coefficient are present in a wave function, this corresponds to high single-orbital entropies for orbitals present in these determinants.
Therefore, we recalculated the H8 model with a CAS(8,8) to ensure that all strongly correlated orbitals are included in the active space.
\begin{figure}[h!t]
  \includegraphics[width=\textwidth]{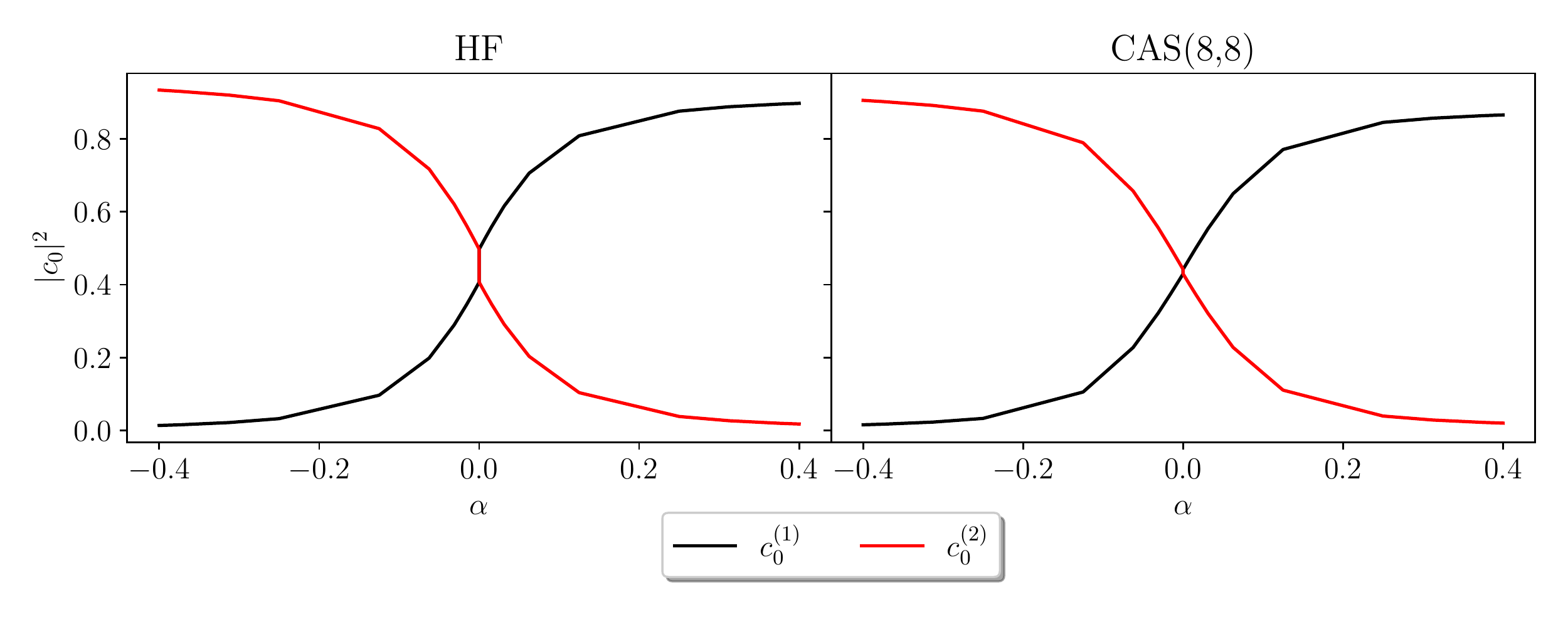}
  \caption{Square of the H8-model CI coefficient ($|c_0|^2$) of $\Phi_1$ ($c_0^{(1)}$) and $\Phi_2$ ($c_0^{(2)}$) of Eqs. (\ref{H8_1}) and (\ref{H8_2}) for different values of $\alpha$. Both expansion coefficients were obtained for determinants constructed from different molecular orbital bases: (1) HF orbitals (left) and (2) CAS(8,8) orbitals (right).}
  \label{referenceCoeff_h8_ex}
\end{figure}
Figure \ref{referenceCoeff_h8_ex} depicts the CI coefficients for the two most dominant reference determinants
given in Eqs. \eqref{H8_1} and \eqref{H8_2} from a CAS(8 $e^-$,$(1a_{g})(1b_{2u})(1b_{3u})(2a_{g})(1b_{1g})(2b_{2u})(2b_{3u})(2b_{1g})$) calculation (see Figure \ref{figOrbEntanglement}).
Since the CAS(8,8) contains 4896 more determinants than the CAS(2,2) and the weight of the most dominant determinant is less than $0.9$, this means that other determinants in the wave function also become important.
In general the CI coefficients of the CAS(8,8) show the same behavior as for CAS(2,2) so that they become degenerate for $\alpha = 0.0$. \\
We performed the analysis of the H8 model with CAS(8,8) in analogy to the analysis for CAS(2,2).
Accordingly, Figure \ref{E_error_h8_ex} shows the absolute energy error with respect to FCI.
From the absolute energy error we conclude that CCSD yields energies with a smaller energy error than TCCSD for the SR-dominated region ($\alpha < -0.09, \alpha > 0.04$).
In contrast to the smaller CAS, TCCSD provides energies with a smaller error than CCSD for the MR region, based on HF orbitals.
However, for the calculations based on CAS(8,8) optimized orbitals, CCSD yields better energies than TCCSD for compressed configurations and for stretched configurations (with $\alpha > 0.12$), whereas for configurations with $0 < \alpha < 0.12$ TCCSD provides energies with a smaller energy error.
Since neither CCSD, nor TCCSD are variational methods, the spikes in these diagrams correspond to sign changes in the error. \\
To gain more insight into the quality of the wave function, Figure \ref{Overlap_h8_ex} shows the overlap error of the wave functions with respect to FCI.
Contrasting the energy error, TCCSD yields an improved wave function compared to CCSD for $-0.23 < \alpha < 0.24$.
It is not surprising that a larger active space yields better wave function quality, since the coefficients which correspond to singly and doubly excited determinants were optimized with respect to highly excited determinants in the active space.
But for configurations with nearly degenerate determinants, the overlap error is still above $0.02$, which would not be the case for a \textit{genuine} MR method.
As expected, the CCSD wave functions feature larger overlaps with the exact reference for SR-dominated regions.

\section{Conclusions}
\label{sec:conclusion}
We presented an analysis of the tailored coupled cluster model TCCSD, which
combines coefficients from a variational complete active space approach with a projective CCSD \textit{ansatz} for the wave function.
We compared energies and wave functions delivered by TCCSD with the
exact results from FCI calculations for well-studied multi-hydrogen
configurations that resemble different correlation regimes by changing a single structural parameter.
We also compared TCCSD with the single-reference CCSD model to understand the benefits and drawbacks of TCCSD
in direct comparison to its SR reference.
As orbital bases we inspected HF as well as CAS-optimized orbitals. \\
One issue of the \textit{tailored} CC approach is the lack of size consistency, which is clear from the evaluation of the P4 energy error based on CASSCF orbitals.
Unlike CCSD, which is size-consistent, the TCCSD results do not converge to the FCI reference in the dissociation limit.
Nevertheless, the overlap error decreases even though the energy error becomes almost constant at larger values of $\alpha$ for the P4 model.
However, in general, CCSD provides smaller overlap and energy errors for the dissociated configurations.\\
Due to the use of CAS amplitudes, TCCSD should capture static electron correlation in molecular structures
which present (almost) degenerate determinants.
For such configurations, TCCSD and CCSD results show a high dependency on the underlying orbital basis.
While for HF orbitals, CCSD shows almost the same or even smaller errors compared to TCCSD,
this changes for CASSCF orbitals,
where the TCCSD energy and overlap errors become smaller than the corresponding ones for CCSD.
Especially for the degenerate case of the H8 model with a CAS(8,8), TCCSD decreases the overlap error. \\
Since the active space amplitudes must not relax during the external amplitude optimization, the quality of TCCSD depends strictly on the choice of the active space.
Therefore, we exploited an entanglement-entropy analysis of all orbitals to identify the orbitals with largest entanglement,
which are the strongly correlated ones to be included in the active space \cite{Boguslawski2012,stei16}.
In general, increased active orbital space sizes improved the TCCSD wave function, i.e.,
the overlap error decreased to a maximum of $0.02$.
Despite increasing the active space to half of the total number of orbitals in the H8 model in the DZ basis, TCCSD still showed a large overlap error in the MR region. \\
Finally, we investigated the dependence on a reference determinant.
Since none of the structure-dependent TCCSD results turns out to be smooth and continuous for the different references determinants,
it confirms and highlights the dependence of a single-reference wave functions. As a result,
the TCCSD method delivers a symmetry-broken wave function.
However, TCCSD can yield a wave function with a smaller overlap error than CCSD for systems with a small but non-negligible MR character, if the active space includes all strongly entangled orbitals.
Hence, it can be used to recover dynamic correlation for CAS-type methods, if the active space is balanced---for which orbital
entanglement in an automated scheme may be exploited \cite{stei16,stei17,Stein2019}---even if this requires large active spaces.\\
We therefore find supporting evidence that TCCSD is an easy implementable approach that improves on the CCSD \textit{ansatz} for strongly correlated systems with comparable computational scaling. Potential drawbacks are (i) missing size-consistency of the wave function, (ii) necessity of a potentially large active space (owing to many strongly entangled orbitals, which eventually will require modern CAS approaches such as DMRG [as in Ref. \cite{Veis2016}] and FCIQMC [as in Ref. \cite{Vitale2020}] and to decrease the size-consistency error), (iii) requirement of CASSCF orbitals, (iv) that systems with a truly strong MR character may not be described properly, and (v) strong dependence on the reference determinant. As we have shown, the determinant with the largest weight in the active space should be employed as reference to reduce potential errors caused by the symmetry broken wave function. \\
To further improve on the accuracy of TCCSD, triples from the active space could be incorporated in the optimization of the external amplitudes. Additionally, it should be extended by approaches that incorporate dynamic correlation effects in the orbitals of the active space.
For instance, this may be accomplished by approaches that are already applied to the CAS-type step such as short-range dynamic correlation in the CAS orbital optimization
\cite{Savin1997_srDFT, Fromager2007, Hedegard2015_DMRG-srDFT} and transcorrelation \cite{Alavi2019_Transcorrelated-Molecules,Baiardi2020}.
Work along these lines is currently in progress in our laboratory.

\section*{Acknowledgement}
This work was financially supported by the Swiss National Science Foundation (project no.\ 200021\_182400).
L.F. acknowledges a Schr\"odinger fellowship (J 3935-N34) of the Austrian Science Foundation (FWF).
We are grateful to Dr. Christopher Stein for helpful comments on the manuscript.

\section*{Data Availability Statement}
The data that support the findings of this study are available from the corresponding author upon reasonable request.
Moreover, see supplementary material in the final publication for figures of orbital bases, potential energy curves, and coupled cluster diagnostics.

%%\bibliographystyle{jcp}
%\bibliography{tcc}
\providecommand{\latin}[1]{#1}
\makeatletter
\providecommand{\doi}
  {\begingroup\let\do\@makeother\dospecials
  \catcode`\{=1 \catcode`\}=2 \doi@aux}
\providecommand{\doi@aux}[1]{\endgroup\texttt{#1}}
\makeatother
\providecommand*\mcitethebibliography{\thebibliography}
\csname @ifundefined\endcsname{endmcitethebibliography}
  {\let\endmcitethebibliography\endthebibliography}{}

\end{document}